\documentclass[fleqn,usenatbib]{mnras}

\usepackage{newtxtext,newtxmath}

\usepackage[T1]{fontenc}
\usepackage{ae,aecompl}

\usepackage{graphicx}	
\usepackage{amsmath}	
\usepackage{amssymb}	

\title[The dust and cold gas content of local star forming galaxies]{The dust and cold gas content of local star forming galaxies}

\author[P.Popesso et al.]{
P. Popesso,$^{1}$\thanks{E-mail: paola.popesso@tum.de}
A. Concas,$^{2}$
L. Morselli,$^{3,4}$
G. Rodighiero,$^{3,4}$
A. Enia,$^{3,4}$
S. Quai$^{5}$
\\
$^{1}$Excellence Cluster Universe, Boltzmannstrasse 2, 85748, Garching bei M\"unchen, Germany\\
$^{2}$Kavli Institute for Cosmology, University of Cambridge, Madingley Road, Cambridge, UK\\
$^{3}$Universit\'a degli studi di Padova, vicolo dell'Osservatorio, Padova, Italy\\
$^{4}$Osservatorio Astronomico di Padova, vicolo dell'Osservatorio, Padova, Italy\\
$^{5}$University of Victoria, 3800 Finnerty Road, Victoria, Canada\\
}

\date{Accepted XXX. Received YYY; in original form ZZZ}

\pubyear{2015}

\begin{document}
\label{firstpage}
\pagerange{\pageref{firstpage}--\pageref{lastpage}}
\maketitle

\begin{abstract}
We use dust masses ($M_{dust}$) derived from far-infrared data and molecular gas masses ($M_{mol}$) based on CO luminosity, to calibrate proxies based on a combination of the galaxy Balmer decrement, disk inclination and gas metallicity. We use such proxies to estimate $M_{dust}$ and $M_{mol}$ in the local SDSS sample of star forming galaxies (SFGs). We study the distribution of $M_{dust}$ and $M_{mol}$ along and across the Main Sequence (MS) of SFGs. 
We find that $M_{dust}$ and $M_{mol}$ increase rapidly along the MS with increasing stellar mass ($M_*$), and more marginally across the MS with increasing SFR (or distance from the relation). The dependence on $M_*$ is sub-linear for both $M_{dust}$ and $M_{mol}$. Thus, the fraction of dust ($f_{dust}$) and molecular gas mass ($f_{mol}$) decreases monotonically towards large $M_*$. The star formation efficiency (SFE, inverse of the molecular gas depletion time) depends strongly on the distance from the MS and it is constant along the MS. As nearly all galaxies in the sample are central galaxies, we estimate the dependence of $f_{dust}$ and $f_{gas}$ on the host halo mass and find a tight anti-correlation. As the region where the MS is bending is numerically dominated by massive halos, we conclude that the bending of the MS is due to a lower availability of molecular gas mass in massive halos rather than a lower efficiency in forming stars.

\end{abstract}


\begin{keywords}
galaxies: evolution -- galaxies: ISM -- galaxies: star formation
\end{keywords}



\section{Introduction}

The interplay between cold gas, dust and star formation in galaxies plays a key role in the studies of galaxy evolution. Star formation occurs within dense, massive, and cold giant molecular clouds. These are the sites where atomic hydrogen is transformed in molecular hydrogen, while the dust grains act as catalyst of the transformation and shield the site from the UV emission of massive young stars. For this reason, atomic, molecular and dust mass appear to be correlated quantities and are often used to derive one another in the study of the star formation process in galaxies. Nevertheless, measuring molecular gas ($M_{mol}$) and dust ($M_{dust}$) masses in galaxies is a challenging task.

The molecular phase (H$_2$) is not directly observable \citep[see][for a review]{kennicutt_evans_12}. Therefore, the luminosity due to the $J=1-0$ transition of the carbon monoxide molecule (CO) is more commonly adopted as a proxy of the H$_2$ gas mass
\citep[e.g.][]{saintonge11, kennicutt_evans_12,bolatto15, saintonge17}. Nevertheless, large CO surveys of galaxies are extremely time consuming. The two largest efforts so far have been the Five College Radio Astronomical Observatory Extragalactic CO Survey \citep[FCRAO,][]{young95}, which measured the CO(1-0) line in 300 nearby galaxies, and the
IRAM-30m reference CO survey for galaxy evolution studies \citep[xCOLD-GASS,][]{saintonge11,saintonge17}, which measured fluxes in the CO(1-0) line for a purely stellar mass selected sample of $\sim$ 500 galaxies at $0.025<z<0.05$ and stellar masses ($M_*$) larger than $10^{9}$ $M_{\odot}$. At higher redshift typical galaxies are now being observed out to $z\sim 0.5$  \citep[e.g.,EGNoG;][]{bauermeister13}, and further out to z$\sim$ 2-3 \citep[e.g.][]{tacconi13, scoville16, scoville17}, but samples are very small in number at fixed redshift window.  With  the Atacama Large Millimeter/submillimeter Array (ALMA) the estimates of $M_{mol}$ will surely increase at low and high-redshift. However, the small instantaneous field-of-view of ALMA does not make this instrument suitable for for wide area galaxy surveys.

In comparison to $M_{mol}$, $M_{dust}$ is relatively easier to measure as dust grains emit mainly in the far infrared (FIR) and sub-millimeter (sub-mm) wavelength range. Early studies of dust content and emission have been done both from space \citep[e.g. IRAS and ISO,see][respectively]{neugebauer84,kessler96} and from the ground \citep[e.g. SCUBA and MAMBO, see][respectively]{holland99,greve08}.

Recent missions in the mid-infrared \citep[e.g. WISE and MIPS,][respectively]{wright10,rieke04}, FIR \citep[e.g. HAKARI and Herschel,][respectively]{hakari,herschel}, and sub-mm \citep[e.g. Planck,][]{planck} have revolutionized the field \citep[e.g.][]{lutz14}. \cite{cortese14} show that as long as the observed SED extends to at least 160-200 $\mu$m in the rest frame, $M_{\rm dust}$ can be recovered with a $>3\sigma$ significance and without significant systematic errors. However, mid and FIR instruments as MIPS on-board \textit{Spitzer}, and PACS and SPIRE on-board \textit{Herschel} are no longer available. Furthermore, the existing FIR data are either deep but limited to a small number of objects \citep[the Herschel Reference Survey, $\sim$ 300 objects,][]{boselli10} or very shallow, sampling mainly the  dustiest objects over large areas 
\citep[H-ATLAS survey over $\sim 660$ deg$^2$,][]{valiante16,2017ApJS..233...26S}.

These methods are powerful but they have the important drawback that do not really allow a proper statistical treatment of the distribution of the dust and molecular mass content as a function of galaxy properties, for example along and across the Main Sequence of star forming galaxies \citep[e.g.][]{noeske07,rodighiero11,popesso19a}. \cite{saintonge11,saintonge17} attempt such analysis exploiting data from the xCOLD-GASS survey. However, the limited number of galaxies at high stellar masses (less than 100) and hosted by massive dark matter halos, prevents a proper statistical analysis of the fraction of molecular gas versus $M_*$ and dark matter halo mass ($M_{halo}$). At higher redshift, \cite{scoville16,scoville17} provide a similar attempt with $\sim$ 1000 galaxies but over a much wider redshift window from z$\sim$1 to z$\sim$4. Similarly, the PHIBBS sample \citep{tacconi13, genzel15} covers the same redshift range with a lower number of objects. 

More recently, \cite{2019MNRAS.486L..91C} propose an alternative method for measuring $M_{mol}$ in galaxies by exploiting dust \emph{absorption} rather than \emph{emission}. They find a tight correlation between the Balmer Decrement (BD, $H\alpha/H\beta$), corrected for the galactic disk inclination, the CO luminosity ($L_{CO}$) and the molecular mass ($M_{mol}$) with a scatter of $\sim$0.3 dex. The huge advantage of this method is that the BD is available for statistically significant samples of star forming galaxies (SFGs), e.g. the SDSS spectroscopic sample in the local Universe. At higher redshift, similar samples will be soon available in huge spectroscopic campaigns with MOONS, PFS and 4MOST. Thus, the use of the BD as a proxy of $M_{mol}$ and $M_{dust}$ opens up the possibility of studying their distribution in galaxies across a variety of system properties and over a large redshift range, where a complete ALMA follow-up will never be possible.

In this paper, we extend the analysis proposed in \cite{2019MNRAS.486L..91C} on the correlation between the BD and $M_{mol}$. In addition, we present a correlation between the BD and $M_{dust}$. We explore, in particular, the role of the galaxy metallicity as possible source of scatter in both the correlations and show the applicability of the BD as proxy of $M_{mol}$ and $M_{dust}$ in a sample of local SFGs drawn from the SDSS spectroscopic sample.

The paper is structured as follows. Section 2 illustrates the dataset used in the analysis. Section 3 describes the calibration of the proxies based on BD, inclination and metallicity for $M_{dust}$ and $M_{mol}$, respectively. Sections 4 shows how the correlations can be used to study the distribution of dust and molecular mass in the local SFGs population. Section 5 shows the relation between the fraction of $M_{dust}$ and $M_{mol}$ with respect to $M_{halo}$. Section 6 summarizes our results. We  assume a $\Lambda$CDM  cosmology  with  $\Omega_M=0.3$, $\Omega_{\Lambda}=0.7$ and $H_0=70$ \textit{km/s/Mpc}, and a Chabrier IMF throughout the paper.

\begin{figure}
    \centering
    \includegraphics[width=\columnwidth]{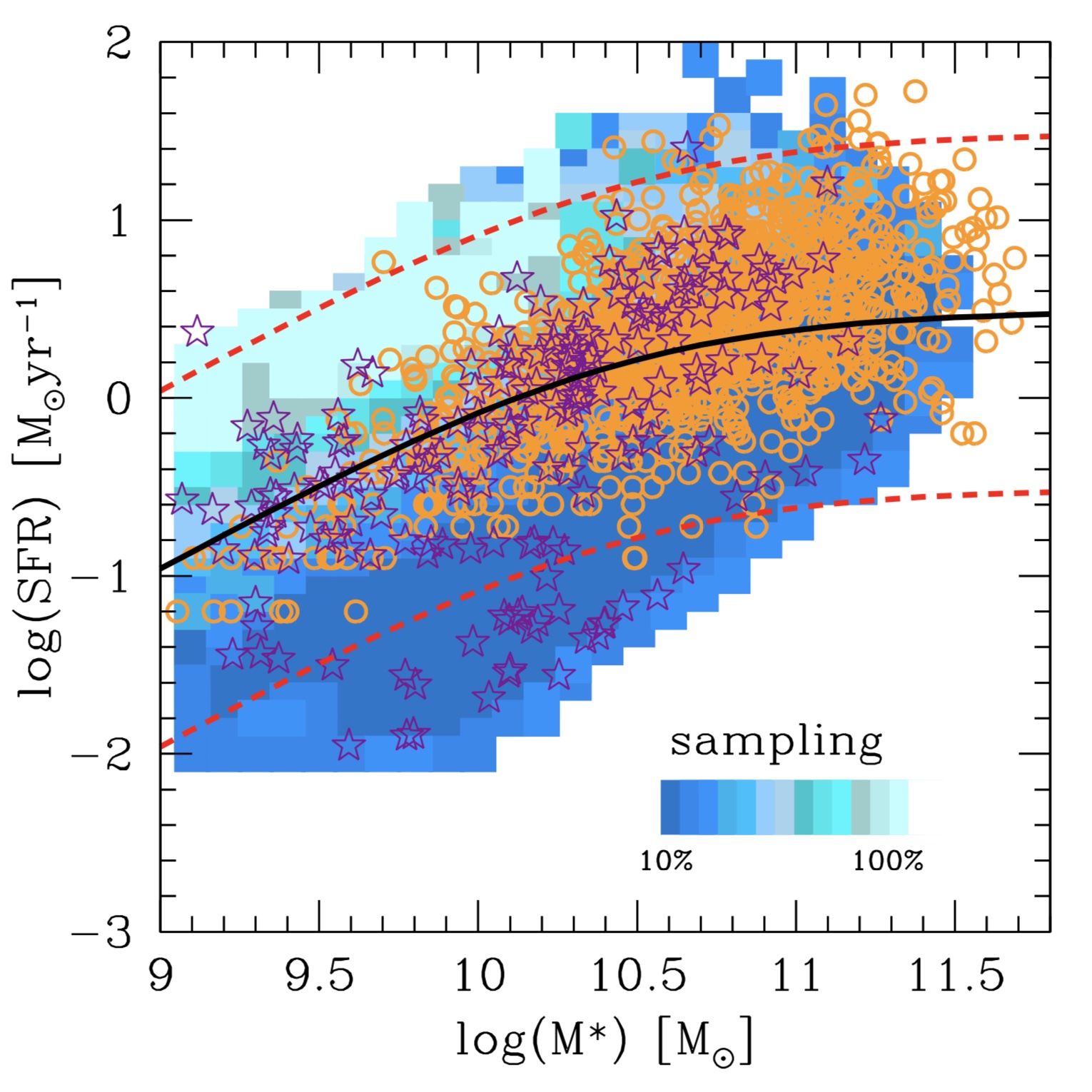}
    \caption{Distribution and sampling rate of the SF galaxy reference sample with respect to the underlying SF galaxy population in the MS region (shaded region). The color-code of the shaded region reflect the sampling rate of the SF galaxy reference sample. The orange empty points indicate the location of the dust mass calibration sample, while the purple stars indicate the location of the molecular gas mass calibration sample. The black solid line marks the MS of \protect\cite{popesso19a}, while the dashed red lines are located 3$\sigma$ above/below the relation.}
    \label{sample}
\end{figure}

\section{Dataset}
The dataset used in this analysis is composed of a local star forming galaxy reference sample and two calibration samples. The former is used to study the distribution of dust and molecular gas across and along the MS of SFGs. The latter two are sub-samples of the local star forming galaxy reference sample with available $M_{dust}$ and $M_{mol}$ estimates, respectively. They are built to cover the same parameter space and redshift window of the reference sample, and to calibrate the proxies of dust and molecular gas mass, based on the BD, disk inclination and galaxy metallicity of the star forming galaxy reference sample.

\subsection{The local star forming galaxy reference sample}
The sample of local SFGs is drawn from the GALEX-SDSS-WISE Legacy Catalog \citep[GSWLC][]{salim16}. GSWLC is obtained by cross-matching the SDSS spectroscopic catalog with the GALEX UV and WISE databases, and also provides SDSS and 2MASS photometric information. It contains galaxies within the GALEX footprint, regardless of a UV detection, and is made up of ~700,000 galaxies with SDSS redshift between $0.01<z<0.30$. We use, in particular, the subsample with medium-deep GALEX observations of $\sim$ 1500s exposure (GSWLC-M), which covers 49\% of the SDSS area. This decision was taken to exploit relatively deep UV observations and large statistics. GSWLC utilizes WISE observations at 22$\mu$m  (WISE channel  W4) to determine SFRs independently of the UV/optical SED fitting. In particular, the mid-IR SFRs in GSWLC are estimated from the total IR luminosity (L$_{IR}$, 8-1000 $\mu$m) by interpolating the luminosity-dependent IR template of \cite{chary01} that matches the 22 $\mu$m flux. For galaxies undetected in WISE, the SFR and dust attenuation are estimated via SED fitting from the UV to the near-IR data. The SED fitting is performed using the state-of-the-art UV/optical SED fitting technique code CIGALE (Noll et al. 2009). The average $5\sigma$ depth in the W4 channel is 5.4 mJy.

The catalog is matched to the MPA-JHU database of galaxy properties\footnote{https://wwwmpa.mpa-garching.mpg.de/SDSS/DR7/} to retrieve stellar masses, $H\alpha$ and $H\beta$ fluxes, the BPT classification and gas metallicity \citep{kauffmann03, brinchmann04,tremonti04,salim07}. When the H$\alpha$ based SFR is available in the MPA-JHU catalog, we use this estimate to replace the SED fitting based SFR of the GSWLC sample. The match between the GSWLC sample and the MPA-JHU catalog leads to the selection of $\sim$350000 galaxies. However, in order to use the BD as a measure of dust attenuation and so as a proxy of the dust mass and the molecular gas mass, the $H\alpha$ and $H\beta$ emission must trace only star formation. Indeed, any AGN contamination would bias the the $H\alpha$ and $H\beta$ fluxes and so their ratio, as found in \cite{2019MNRAS.486L..91C}. For this reason, we further isolate SFGs on the basis of the BPT  classification provided in the MPA-JHU catalogs. We further limit the sample to galaxies at $z < 0.1$ in order to have a 80\% and 50\% spectroscopic completeness above stellar masses of $10^{10}$ and $10^9$ $M_{\odot}$, respectively. We exclude dwarf galaxies below $10^9$, for which the dust/molecular gas properties might significantly differ from more massive star forming galaxies \citep{2019A&A...626A..23C}.

Prior to the estimate of the BD, we multiply the uncertainties of the $H\alpha$ and $H\beta$ fluxes by a correction factor (f=2.473 and f=1.882 for the $H\alpha$ and $H\beta$ line, respectively), to take into account continuum subtraction errors \citep[following][]{brinchmann04}. As shown in \cite{2019MNRAS.486L..91C}, the BD is a good proxy of $M_{mol}$ once corrected for the galaxy disk inclination. To retrieve an accurate value of the galaxy disk inclination we use the measures provided by the bulge-disk decomposition catalogue of \cite{2011ApJS..196...11S}, available for nearly all systems (98\%) of the star forming galaxy sample.

The final sample comprises $\sim$ 150.000 star forming systems, which we will refer to as the "star forming galaxy reference sample" throughout the paper. As explained in \cite{brinchmann04}, such SF galaxy sample might induce several biases, because the BPT classification requires a SNR$>3$ in several emission lines (H$\alpha$, H$\beta$, NII and [OIII]). The emission line galaxy population selected in this way might not be fully representative of the star forming galaxy population in the MS region. To check this aspect, we estimate the sampling rate of the star forming galaxy reference sample with respect to the parent GSWLC sample in the same redshift window. The result in shown in Fig. \ref{sample}. The star forming galaxy reference sample populates the whole MS region at all stellar masses within 3$\sigma$ from the MS relation, taken from \cite{popesso19a} and estimated exploiting the GSWLC catalog. However, the sampling rate is not uniform across and along the MS. The star forming galaxy reference sample captures a very high percentage ($\sim$50\%) of the galaxies above the MS, mainly in the starburst region and in the stellar mass range $10^9-10^{10.4}$ $M_{\odot}$. The sampling rate remains quite constant and uniform (17-22\%) elsewhere. To correct for this bias we define for each galaxy a weight proportional to the mean sampling rate estimated in a region of 0.2 dex $\times$0.2 dex in $M_*$ and SFR around each galaxy. These weights are considered when fitting the data in the following Sections.

\subsection{The calibration samples}

\subsubsection{Molecular masses}
As in \cite{2019MNRAS.486L..91C}, the sample of galaxies with available CO luminositiy ($L_{CO}$) is taken from the extended CO Legacy Database for GASS survey \citep[xCOLD GASS][]{saintonge17}, designed to provide a picture of molecular gas across the local galaxy population. The sample contains 532 galaxies and is obtained by the combination of the original COLD GASS survey, targeting galaxies at $0.025<z<0.050$ and $M_*>10^{10}$ $M_{\odot}$ \citep{saintonge11}, with the COLD GASS-low survey, extended for sources with $10^9M_{\odot}<M_{*}<10^{10}M_{\odot}$, \citep{saintonge17}. 

A percentage of 63\% of the sample (333 objects) is CO detected, while for few of the remaining galaxies upper limits are provided \citep[see][]{saintonge11, saintonge17}. $M_{mol}$ is derived from $L_{CO}$ through the $\alpha_{CO}$ factor conversion, calibrated in \cite{accurso17} and dependent on galaxy metallicity. The 532 xCOLD GASS galaxies are cross-matched with the star forming galaxy reference sample to retrieve SFRs, $M_*$, $H\alpha$ and $H\beta$ fluxes, disk inclination angles and metallicities.  This leads to a sample of 201 star forming systems: 177 galaxies with detected CO emission and 24 upper limits. The upper limits are combined in a single point of average $L_{CO}$ and $M_{mol}$.

To complement xCOLD GASS in the starburst region, where the Balmer Decrement might saturate due to the high level of dust content, we include the Herschel SPIRE-selected sample of \cite{bertemes18} in the SDSS Stripe 82 area. The shallow Herschel-Stripe 82 survey leads to the selection of the dustiest local objects. \cite{bertemes18} provide a measure of $L_{CO}$ and $M_{mol}$ for a subsample of 78 WISE and SPIRE simultaneously detected systems with ALMA follow-up observations of CO. \cite{bertemes18} derive $M_{mol}$ from $L_{CO}$ through the $\alpha_{CO}$ factor conversion based on the metallicity calibration of \cite{2013ApJ...773...68G} and \cite{tacconi18}. For consistency with the xCOLD GASS data, we re-derive $M_{mol}$ with the $\alpha_{CO}$ factor conversion of \cite{accurso17}. For this purpose we use the metallicity provided by \cite{bertemes18}, which is based on the \cite{2004MNRAS.348L..59P} calibration, as in \cite{accurso17}. Nevertheless, we point out that the two $\alpha_{CO}$ calibrations lead to consistent results, as already stated in \cite{bertemes18}. As already shown in \cite{2019MNRAS.486L..91C}, the starburst sample follow the BD-$L_{CO}$ and BD-$M_{mol}$ correlations of the xCOLD-GASS sample at the highest $L_{CO}$ and $M_{mol}$.

The sample is cross-matched to the star forming reference galaxy sample to retrieve SFRs, $M_*$, $H\alpha$ and $H\beta$ fluxes, disk inclination and metallicities. The match excludes objects at $z > 0.1$ and leads to a subsample of 53 starburst galaxies.

The combination of the xCOLD-gass and \cite{bertemes18} starburst samples leads to 220 galaxies with CO detection and 24 upper limits. We refer to this sample throughout the paper as "the molecular mass calibration sample". The distribution of these objects in the Log(SFR)-Log($M_*$) plane is shown in Fig. \ref{sample} (purple stars). The sample nicely populates the entire MS region. As for the star forming galaxy reference sample, to take into account possible biases, we estimate the sampling rate of the calibration sample with respect to the parent GSWLC sample in the same redshift window. We define for each galaxy a weight proportional to the mean sampling rate estimated in a region of 0.2 dex $\times$0.2 dex in $M_*$ and SFR around each galaxy. These weights are considered when fitting the data in the following Sections.

\begin{figure*}
\includegraphics[width=\columnwidth]{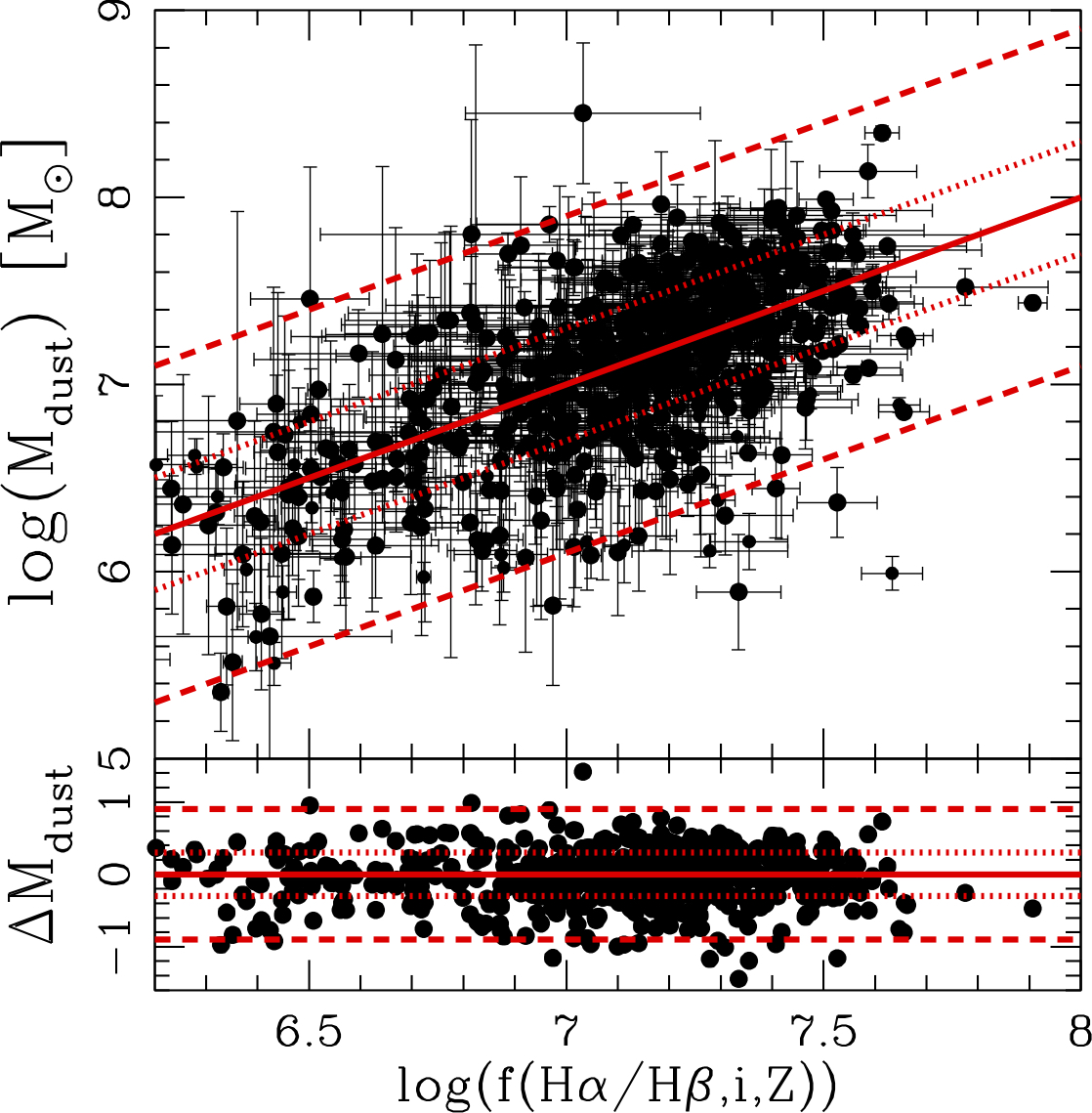}
\includegraphics[width=\columnwidth]{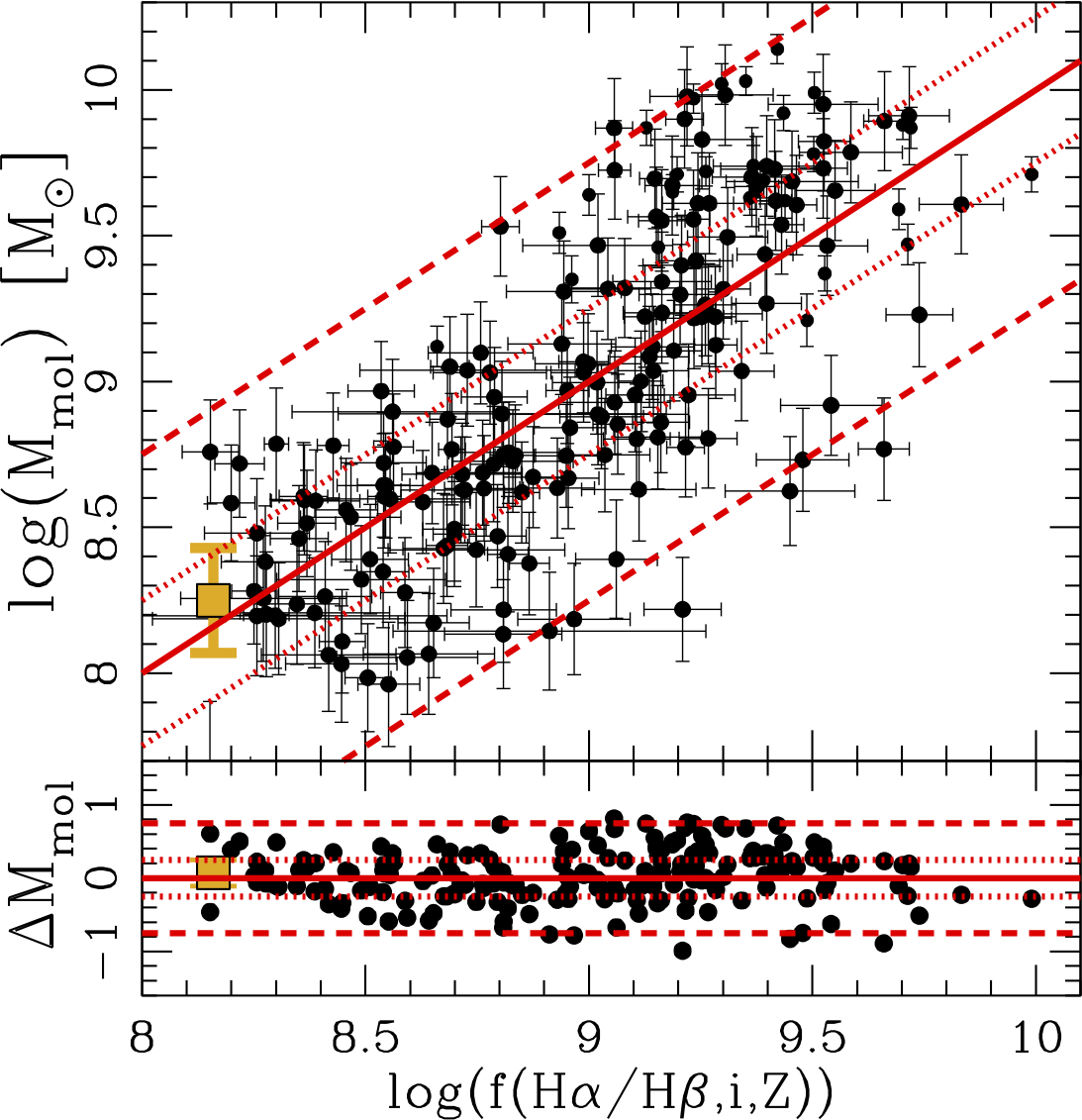}
\caption{{\it{Left panel}}: The upper panel shows the $log(M_{dust})-f_{dust}(BD,i,Z)$ relation. The red solid line shows the 1:1 relation. The dotted line shows the 1$\sigma$ uncertainty, while the dashed line shows the 3$\sigma$ level. The bottom panel shows the residual distribution with respect to the 1:1 relation. The red solid, dotted and dashed lines show the 0 residual level, and the 1 and 3$\sigma$ scatter around the 1:1 line, respectively. {\it{Right panel}}: same as in the left panel for the $log(M_{mol})-f_{mol}(BD,i,Z)$ relation for the molecular gas calibration sample.}
\label{corr}
\end{figure*}

\subsubsection{Dust masses}

The sample of galaxies with available $M_{dust}$ estimates is obtained combining different Herschel-based surveys. The largest fraction of sources comes from the Herschel Astrophysical Terahertz Large Area Survey (the Herschel ATLAS or H-ATLAS) conducted with {\it{Herschel}} PACS and SPIRE. The H-ATLAS is a survey of approximately 660 deg$^2$ of sky in five photometric bands: 100,160, 250, 350 and 500 $\mu$m \citep{eales10}. The catalog used in this work is based on the H-ATLAS DR2 \citep{maddox18}. The DR2 covers two H-ATLAS fields at the north and south Galactic poles for a total area of $\sim$500 deg$^2$. 

We use the SPIRE emission at 250, 350 and 500 $\mu$m to estimate $M_{dust}$. This is done following the procedure presented in Dunne et al. (2011), by means of a SED fitting with a modified grey-body law $S \propto \nu^{\beta} B(\nu, T)$ with fixed emissivity index $\beta = 1.5$ in order to retrieve the dust temperature $T_{\rm dust}$, and then dust masses from:
\begin{equation}
M_{\rm dust} = \dfrac{S_{250} D^2 (1+z) K}{k_{250} B(\nu_{250}, T_{\rm dust}) }
\end{equation}
where $k_{250}$ is the dust mass absorption coefficient (0.89 m$^2$ kg$^{-1}$ at 250 $\micron$) and K the {\it K-correction}.

As done in previous section, the H-ATLAS catalog is cross-matched with the star forming galaxy reference sample, to retrieve SFRs,  $M_*$, $H\alpha$ and $H\beta$ fluxes, BPT classifications, disk inclination and metallicities. This leads to a sample of $\sim$1200 far-infrared selected star forming systems. 

To complement the H-ATLAS subsample at the low $M_{dust}$ end (as H-ATLAS maps are relatively shallow) we include in our sample the data from the deeper Herschel Reference Survey \citep[HRS,][]{boselli10}. The HRS sample is volume-limited, containing sources with flux limits in the K-band to minimize the selection effects associated with dust and with young high-mass stars and to introduce a selection in stellar mass. Galaxies in HRS span the whole range of morphological types (ellipticals to late-type spirals) and environments (from the field to the centre of the Virgo Cluster). The estimates of $M_{dust}$ are taken from \cite{cortese14} and are based on SED fitting of the SPIRE and PACS data. The cross-match between the HRS and our star forming galaxy reference samples results in 157 galaxies. The star forming galaxy selection limits also the bias towards dense environment due to the inclusion of the Virgo Cluster galaxy population, which is dominated by early type systems.

Finally, we include the \cite{bertemes18} 53 sources in the SDSS Stripe 82 area that have available $M_{dust}$ estimates, matched to the star forming galaxy reference sample.

In total we collect a sample of $\sim$1400 galaxies with $M_{dust}$ measurements based on {\it{Herschel}} SPIRE data. We refer to this sample as the "dust mass calibration sample". The distribution of the selected objects in the Log(SFR)-Log($M_*$) plane is shown in Fig. \ref{sample} (orange empty circles). The sample covers the entire MS region above stellar masses of $\sim$ $10^{10}$ $M_{\odot}$. Below this threshold, most sources lie near the peak of the distribution, while the starburst region is under-represented. As for the molecular mass calibration sample, we estimate a weight for each galaxy on the basis of the sampling rate with respect to the parent GSWLC sample.

\begin{table}
\begin{center}
\begin{tabular}{ccccc}
    \hline
    \hline \\[-2.5mm]
     & $\alpha$ & $\beta$ & $\gamma$ & $\delta$ \\
  \hline
    \hline \\[-2.5mm]
$M_{dust}$ & -6.11$\pm$0.02 & 1.58$\pm$0.02 & -0.003$\pm$0.002& 1.46$\pm$0.03 \\
$M_{mol}$ & 8.72$\pm$0.03 & 4.29$\pm$0.02 & 0.008$\pm$0.006 & 0.65$\pm$0.01\\
 \hline
   \hline
\end{tabular}
\end{center}
\caption{The table lists the best fit values of the fitting form of Eq. \ref{best_fit} for the dust and the molecular mass.}
\label{fit_param}
\end{table}

\begin{figure*}
\includegraphics[width=\columnwidth]{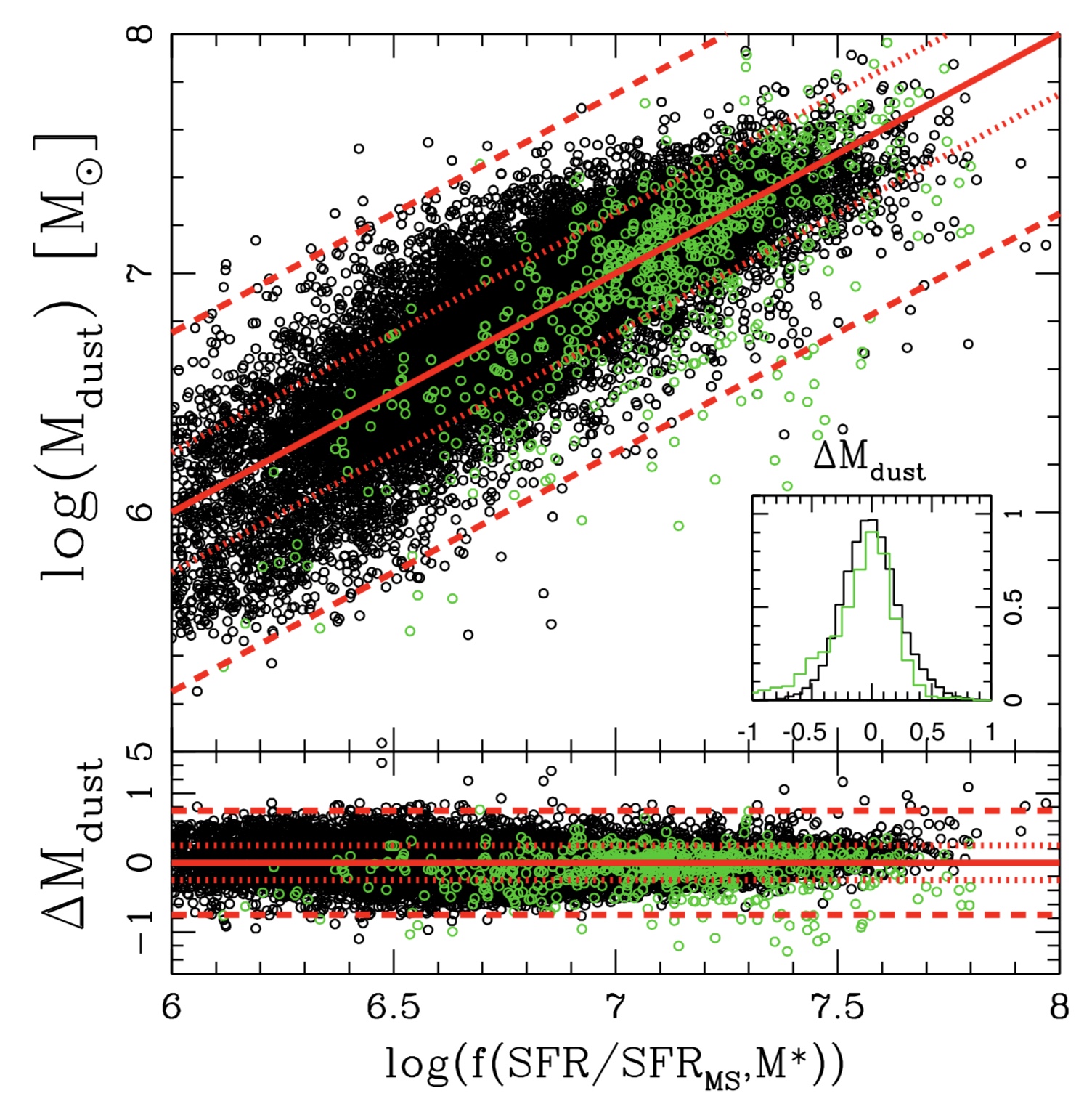}
\includegraphics[width=\columnwidth]{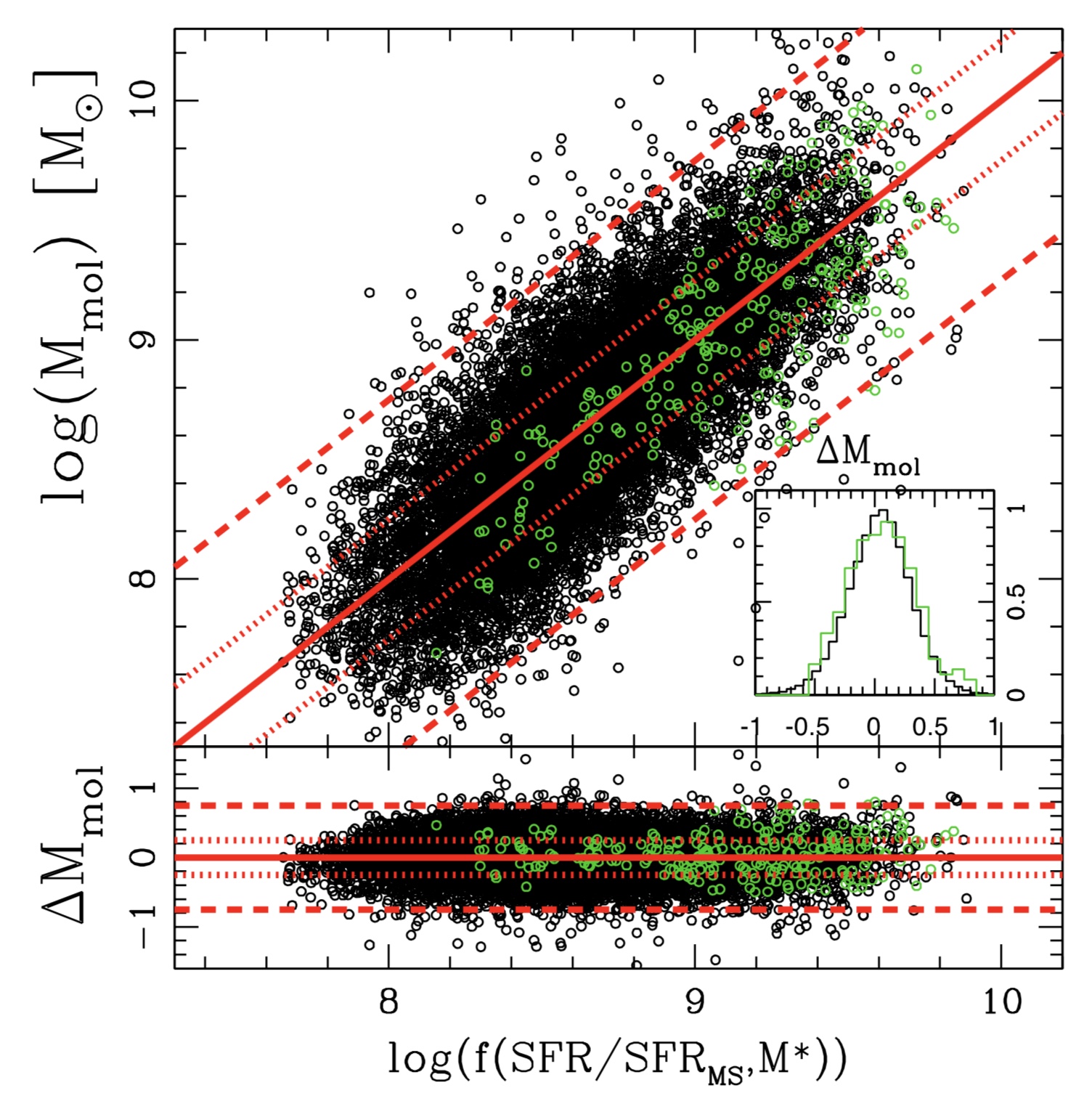}
\caption{{\it{Left panel}}: The upper panel shows how well $M_{dust}$ can be expressed as a function of $SFR/SFR_{MS}$ and $M_*$ according to eq. \ref{function1}. The black points show the star forming galaxy reference sample. The green points shows to the dust mass calibration sample. The red solid, dotted and dashed lines show the 1:1 relation, and the 1 and 3$\sigma$ levels, respectively. The small panel shows the histograms of the residual distributions (black for the SFG reference sample, green for the calibration sample). The bottom panel shows the residual distribution with respect to the 1:1 relation. The red solid, dotted and dashed lines show the 0 residual level, and the 1 and 3$\sigma$ levels, respectively. The points are color-coded as in the upper panel. {\it{Right panel}}: same as in the left panel for $log(M_{mol})$. In these panels the green points show the molecular mass calibration sample.}
\label{check}
\end{figure*}

\section{The proxy for $M_{dust}$ and $M_{mol}$}

As shown in \cite{2019MNRAS.486L..91C}, once corrected for the effect of the disk inclination, the BD is a good proxy of the molecular gas mass. This is because for very inclined galaxies the auto-absorption of the $H\alpha$ and $H\beta$ emission from the galactic disk tends to saturate the BD and enlarge the scatter of the correlation. In addition to the disk inclination, here we also consider the galaxy metallicity. \cite{2019MNRAS.486L..91C} find a poorly significant correlation between the residuals around the $L_{CO}-BD$ relation and the gas metallicity ($Z$). However, $Z$ is recognized has a key parameter of the correlation between dust and cold gas in galaxies from the observational \citep{2007ApJ...663..866D, 2009ApJ...701.1965M} and the theoretical point of view \citep{2013MNRAS.432.2112B}. Thus, we include it when calibrating the proxy for $M_{dust}$ and $M_{mol}$. To find the best proxy, we express $log(M_{dust})$ and $log(M_{mol})$ as a function of BD, inclination and metallicity with the following functional form:

\begin{equation}
f(BD,i,Z)=\alpha+\beta*log(BD-BD_0)+\gamma*(i-i_0)+\delta*Z
\label{best_fit}
\end{equation}
where $i$ is the disk inclination angle, $BD_0$ and $i_0$ are the mean values of the Balmer decrement and inclination in the sample, respectively ($BD_0=0.63$ and $i_0=59$ deg), and $Z=log(12+O/H)$ is the gas metallicity. We fit the $log(M_{dust})-f_{M_{dust}}(BD,i,Z)$, and $log(M_{mol})-f_{M_{mol}}(BD,i,Z)$ relations to retrieve the best fit parameters. The fit is performed by weighting each galaxy as explained in Section. 2.2 in order to correct for any selection bias. Table 1 lists the best fit values and the uncertainties of the 4 free parameters of Eq. \ref{best_fit}, for $M_{dust}$ and $M_{mol}$, respectively. In Fig. \ref{corr} we show the agreement between the proxies estimated as $f_{M_{dust}}(BD,i,Z)$, $f_{M_{mol}}(BD,i,Z)$ and the observed $log(M_{dust})$ and $log(M_{mol})$, respectively. The points scatters around the 1:1 relations with a scatter of 0.24 and 0.22 dex for the $log(M_{dust})-f_{M_{dust}}(BD,i,Z)$ and $log(M_{mol})-f_{M_{mol}}(BD,i,Z)$ relations, respectively.


The scatter of the $log(M_{mol})-f_{M_{mol}}(BD,i,Z)$ relation is slightly smaller than what found in \cite{2019MNRAS.486L..91C}, likely because of the introduction of the gas metallicity as additional parameter in the proxy. The distribution of the residuals for both correlations is shown in the bottom panels of Fig. \ref{corr}. In the right panel of Fig. \ref{corr}, the mean value of the CO upper limits provided by \cite{saintonge17} for the xCOLD-GASS sample is indicated by the gold square. This point lies on the best fit as already found in \cite{2019MNRAS.486L..91C}.

\begin{figure*}
\includegraphics[width=\columnwidth]{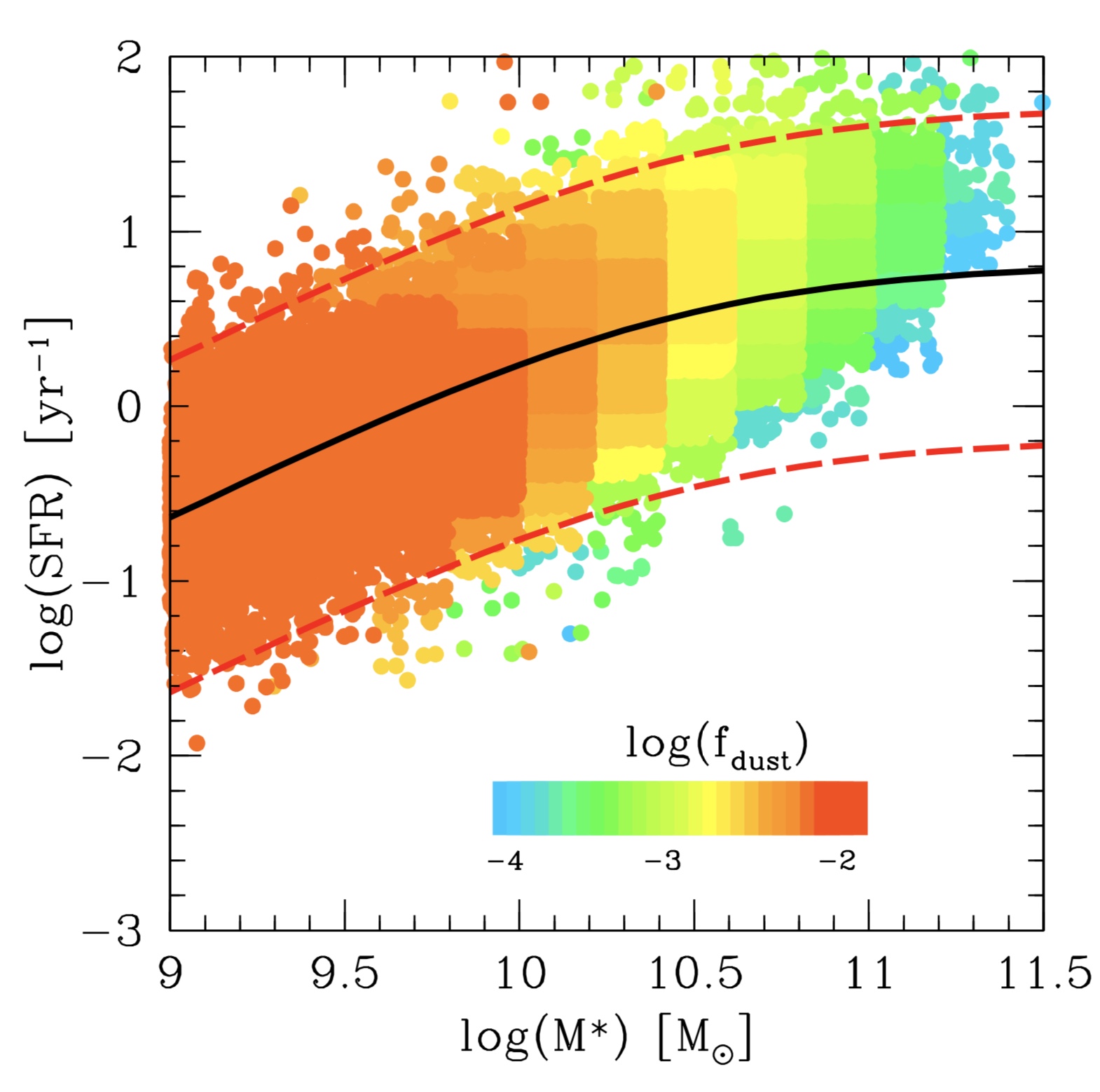}
\includegraphics[width=\columnwidth]{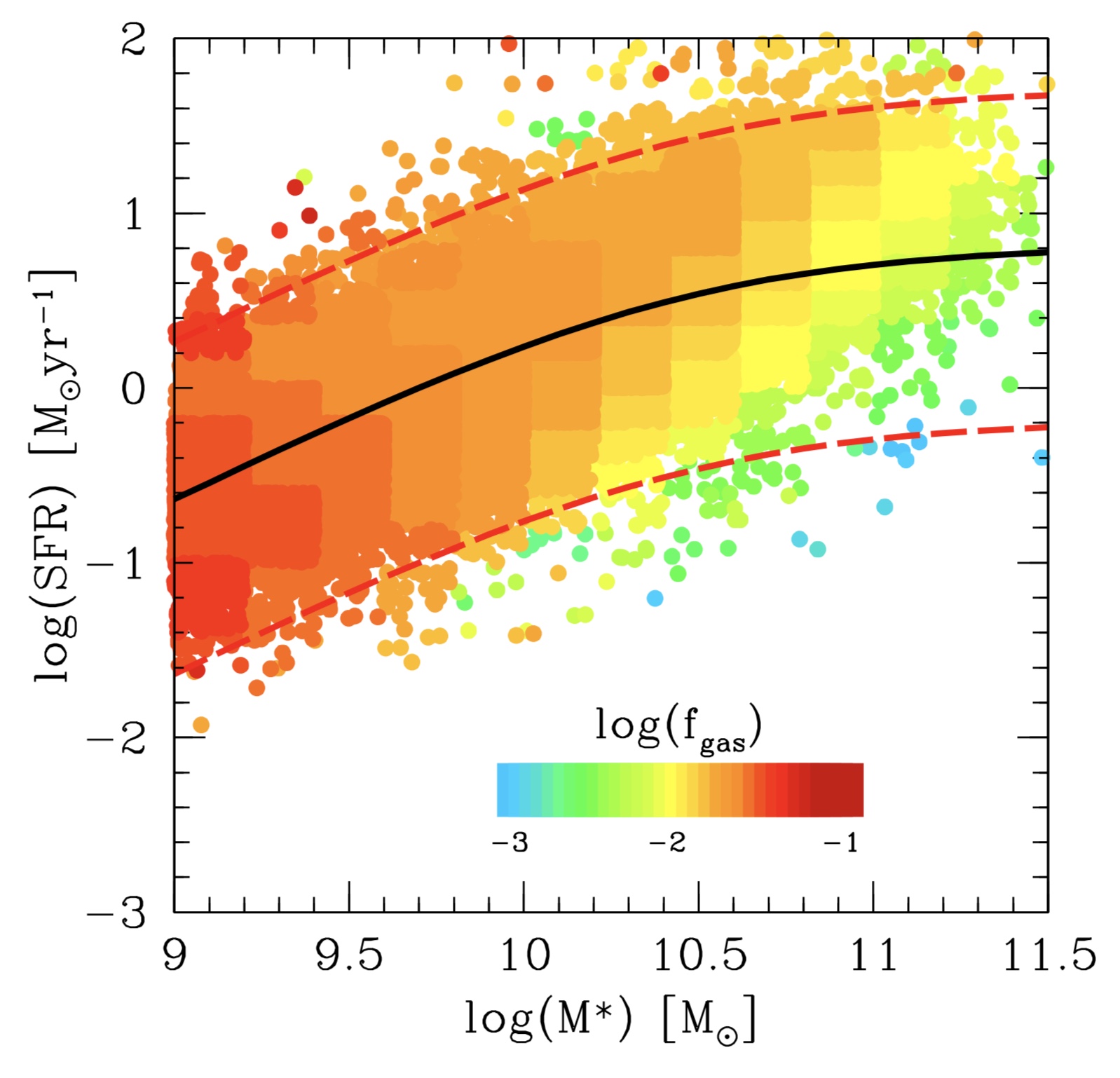}
\caption{{\it{Left panel}}: Distribution of the star forming galaxy reference sample color-coded as a function of the dust mass fraction ($f_{dust}$). The color-code is indicated in the figure {\it{Right panel}}: Distribution of the star forming galaxy reference sample color-coded as a function of the dust mass fraction ($f_{mol}$). The color-code is indicated in the figure}
\label{frac}
\end{figure*}

As a caveat, we point out that the proxies calibrated here for $M_{dust}$ and $M_{mol}$ are applicable only to the SDSS or similar fiber spectroscopy dataset with the same fiber aperture, and redshift window. We do not observe significant dependence of the best fit residuals on the SDSS fiber aperture of 3 arcsec diameter. However, the Balmer decrement within the fiber is an average measure of the relative dust reddening of two fluxes and it might change if larger or smaller apertures are considered. We do not attempt here any fiber aperture correction, because the data do not allow to do so. Spatially resolved maps of larger samples of galaxies, such as CALIFA and MaNGA, will allow to study more in detail the dependence of dust and molecular gas content on the BD within the same physical region. 

\section{The dust and cold gas content along and across the MS}

\begin{table*}
\begin{center}
\begin{tabular}{lcccc|lcccc}
    \hline
    \hline \\[-2.5mm]
\multicolumn{5}{c}{$10^{\alpha}\times(SFR/SFR_{MS})^{\beta}\times{M_*}^{\delta}$}& \multicolumn{5}{c}{$10^{\alpha_1}\times SFR^{\beta_1}\times{M_*}^{\delta_1}$} \\
    \hline
    \hline \\[-2.5mm]
     & $\alpha$ & $\beta$ &  $\delta$ & $\sigma$ & & $\alpha_1$ & $\beta_1$ &  $\delta_1$ & $\sigma$\\
  \hline
    \hline \\[-2.5mm]
$M_{dust}$ & 1.87$\pm$0.01 & 0.12$\pm$0.01 & 0.50$\pm$0.01 & 0.22 & $M_{dust}$ & 0.78$\pm$0.02 &0.04$\pm$0.01 & 0.62$\pm$0.01 & 0.20\\
$M_{mol}$ & 2.99$\pm$0.01 & 0.15$\pm$0.01 & 0.58$\pm$0.02 & 0.21 & $M_{mol}$ & 2.61$\pm$0.01 &0.11$\pm$0.01 &0.61$\pm$0.02 & 0.23 \\
$f_{dust}$ & 2.03$\pm$0.01 & 0.097$\pm$0.01 & -0.51$\pm$0.01 & 0.23 & $f_{dust}$ & 0.86$\pm$0.02 &0.01$\pm$0.01 & -0.39$\pm$0.01 & 0.20\\
$f_{mol}$ & 3.34$\pm$0.01 & 0.10$\pm$0.01 & -0.45$\pm$0.01 & 0.21 & $f_{mol}$ & 2.88$\pm$0.02 &0.05$\pm$0.01 & -0.40$\pm$0.01 & 0.20\\
$SFE$ & -10.25$\pm$0.01 & 0.88$\pm$0.02 & 0.11$\pm$0.01 & 0.14 & $SFE$ & -3.48$\pm$0.02 &0.87$\pm$0.02 & -0.52$\pm$0.01 & 0.20\\
 \hline
   \hline
\end{tabular}
\end{center}
\caption{The table lists the best fit parameters of the fitting forms given in Eq. \ref{function1} and \ref{function2} for $M_{dust}$, $M_{mol}$, $f_{dust}$, $f_{mol}$ and SFE, respectively.}
\label{para}
\end{table*}

In this Section we use the proxies found in the previous analysis to estimate $M_{dust}$ and $M_{mol}$ for all galaxies in the local star forming galaxy reference sample. This allows us to study with unprecedented statistics the distribution of $M_{dust}$ and $M_{mol}$ along and across the MS. To compare our results with others in literature at low \citep{saintonge11,saintonge17} and high redshift \citep{tacconi13,scoville16, scoville17} we parametrize $M_{dust}$ and $M_{mol}$ as a function of $M_*$, the distance from the MS ($SFR/SFR_{MS}$) and SFR. We use as reference the MS of \cite{popesso19a} as it is based on the same star forming galaxy reference sample used here. All fits are performed by taking into account the weights of the star forming galaxy reference sample in order to correct for under-sampling in several regions of the MS. We adopt the following fitting functions:

\begin{equation}
M_{dust}=10^{\alpha}\times(SFR/SFR_{MS})^{\ \beta}\times{{M_*}^{\delta}}
\label{function1}
\end{equation}

\begin{equation}
M_{dust}=10^{\alpha_1}\times SFR^{ \ \beta_1}\times{M_*}^{\delta_1}
\label{function2}
\end{equation}

The same fitting forms are assumed for $M_{mol}$, for the fraction of dust and molecular mass and for the star formation efficiency, defined as $SFE=SFR/M_{mol}$ as in \citet[][ the inverse of the molecular gas depletion time]{2014A&A...562A..30S}. In all cases, a combination of $M_*$ and $SFR/SFR_{MS}$ or SFR can reproduce accurately the dust and molecular gas content of SFGs with a scatter varying from 0.19 to 0.22 dex. The best fit parameters are provided in Table \ref{para} for all the considered variables. The errors are estimated via bootstrapping. Fig. \ref{check} shows an example of the best fit result for the dust (left panel) and molecular mass (right panel). As a further check, we apply the best fitting function also to the calibration samples of dust and molecular mass to check if the best fit can reproduce the observed values (green points in the figure). In both cases, the best fitting functions provide a very good match to the observed dust and molecular masses, with a residual distribution consistent with the reference sample. We observe only a slight asymmetry in the distribution of the residuals for $M_{dust}$, as shown in the left panel of Fig. 3. However, this is not statistically significant. We point out that expressing $M_{dust}$ and $M_{mol}$ as a function of the MS position does not lead to a better fit but it introduces the uncertainties of the MS location. Both dust and molecular masses can be well reproduced as a function of SFR and stellar mass. 

\begin{figure}
\includegraphics[width=\columnwidth]{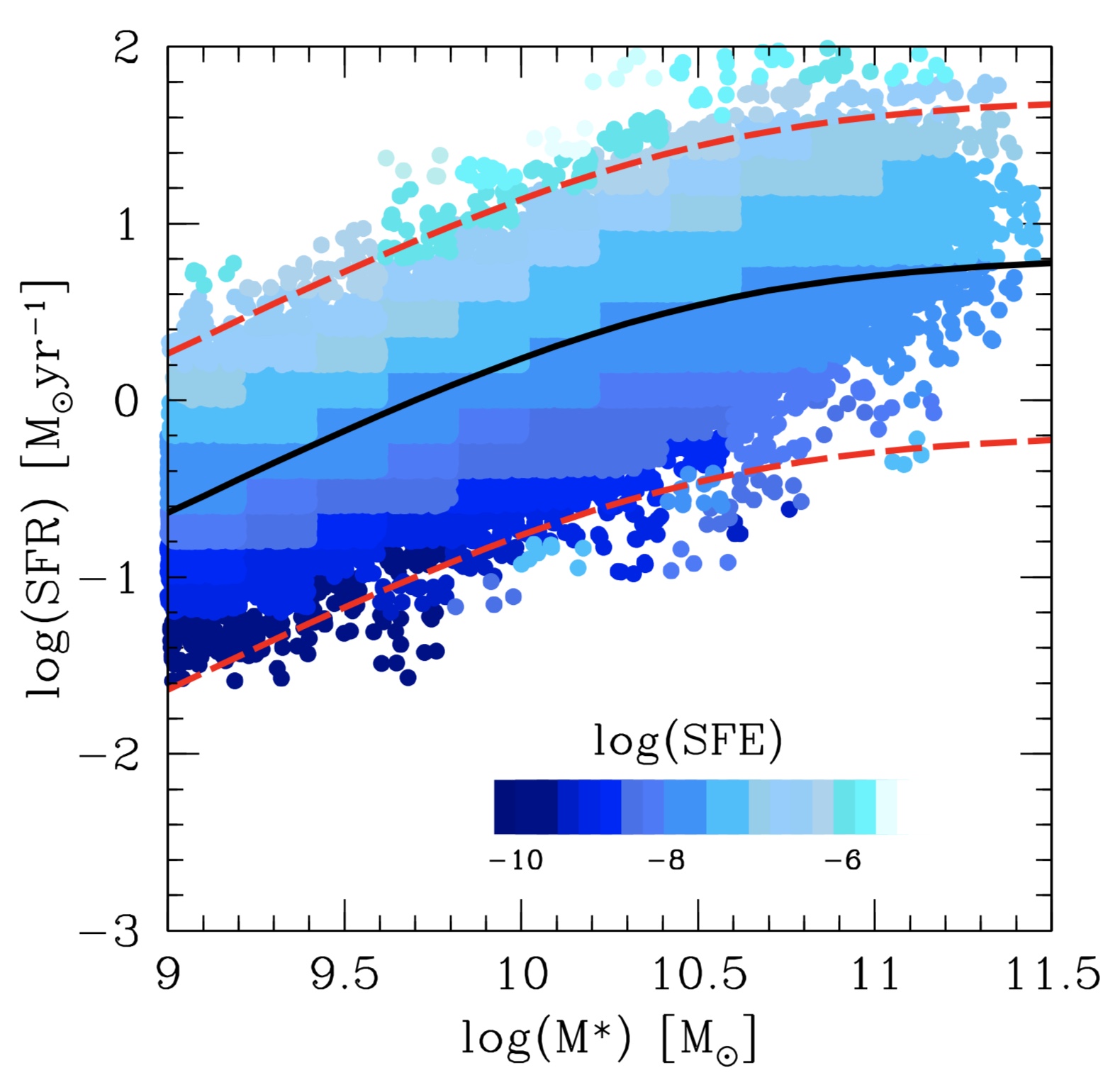}
\caption{Distribution of the star forming galaxy reference sample color-coded as a function of SFE. Each bin is color-coded as a function of the average SFE in the bin.}
\label{sfe}
\end{figure}

The best fit results reveal that both $M_{dust}$ and $M_{mol}$ increase as a function of the $M_*$ more rapidly than with the SFR or $SFR/SFR_{MS}$. $M_{dust}$ exhibits a less significant dependence on the SFR than $M_{mol}$ (see Table \ref{para}). In both cases, the dependence on $M_*$ is sub-linear, which implies that the fractions of dust ($f_{dust}$) and molecular gas masses ($f_{mol}$) are both decreasing along the MS, by almost two orders of magnitude, as clearly visible in Fig. \ref{frac}. The figure shows the MS region color coded as a function of the mean $f_{dust}$ (left panel) and $f_{mol}$ (right panel). It is quite visible that the most significant gradient is along rather than across the relation. Indeed, both $f_{dust}$ and $f_{mol}$ exhibit a poorly significant dependence on SFR or $SFR/SFR_{MS}$, as indicated in Table \ref{para}. The lowest fraction of $M_{dust}$ and $M_{mol}$ is reached where the MS bends towards higher masses. 

Our results are qualitatively in agreement with previous studies, but we point out a quantitative discrepancy. \cite{saintonge11} find very similar results on the basis of the xCOLD-GASS sample. However, they report that the fraction of molecular gas depends mainly on the galaxy specific SFR (sSFR=SFR/$M_*$). Instead, the different dependence on $M_*$ and SFR found here implies that sSFR is not the best proxy. Indeed, when fitting $f_{mol}$ as a function of the sSFR we obtain a decent fit but with a much larger scatter (0.41 dex), with respect to the fitting form of Eq. \ref{function2}. We speculate that the discrepancy of the results might be induced by the limited number of galaxies of the xCOLD-GASS sample with respect to the statistics offered by our star forming galaxy reference sample. An additional source of discrepancy is that \cite{saintonge17} is based on the stacking analysis of the IRAM spectra in several regions of the LogSFR-Log$M_*$ plane. Thus, the discrepancy might also arise from averaging $M_{mol}$ and $f_{mol}$ in bins of SFR and $M_*$  rather than considering the full sample.

\cite{scoville16,scoville17} find, at much higher redshift, that $M_{mol}$ increases in the same way with $M_*$ and $SFR/SFR_{MS}$ as $\propto {SFR/SFR_{MS}}^{0.32}*M_*^{0.3}$. Despite the quantitative discrepancy, the relation of \cite{scoville17} suggests, as our findings, that $f_{mol}$ decreases much faster with the $M_*$ than with SFR. So they also find that sSFR can not be the key parameter in expressing $M_{mol}$ or $f_{mol}$.  The same level of agreement is found also with the results of \cite{2014A&A...562A..30S}.

We also estimate the star formation efficiency and study how this quantity varies along and across the MS. According to the best fits of eq. \ref{function1} and \ref{function2}, SFE is the only parameter, for which the use of the eq. \ref{function1} leads to a smaller scatter (0.14 dex), with respect to eq. \ref{function2} (0.2 dex). In addition, once the distance from the MS is considered, the dependence on the stellar mass becomes negligible, as reported in Table \ref{para}. Fig. \ref{sfe} shows the MS region color coded as a function of the mean SFE. It is quite visible that the location of the MS is characterized by a constant value of SFE, which increases rapidly as a function of the distance from the MS ($SFE\propto SFR/SFR_{MS}^{0.88}$). The value of SFE on the MS is $\sim 1.05\pm0.3 \times10^{-9}$ $yr^{-1}$, which is consistent with the average depletion time ($1/SFE$, e.g. the time needed to consume the available gas mass) estimated for the xCOLD-GASS sample in Saintonge et al. (2013). SFE varies by almost 4 order of magnitudes across the MS over the 3$\sigma$ region around the relation.

This is consistent with previous results in literature, which find that the depletion time depends mainly on the distance from the MS, both at low \citep{saintonge13} and high redshift \citep{tacconi13, genzel15,scoville17}. \cite{2015A&A...575A..74S} find, instead, that SFE decreases at large stellar masses at $z\sim 1$, and it is the cause for the bending of the relation. The discrepancy might arise from the method used to estimate the gas mass. \cite{2015A&A...575A..74S} use the dust-to-gas mass ratio, corrected for the gas metallicity. This is a proxy for the total gas mass ($M_{HI}+M_{H2}$), including atomic and molecular phases. Here, instead, we calibrate our proxy only to provide the mass of the molecular gas phase, similarly to the literature results cited above.

We conclude that the lower fraction of dust and molecular gas mass at the high mass end, rather than a lower efficiency in forming stars, is more likely to explain the bending of the MS. The reduced availability of fuel for the star formation process, would lead, indeed, to a lower SFR in galaxies, consistently with the results obtained by \cite{saintonge17}.

\section{A tight relation with the host halo mass}

\cite{popesso19a} find that the MS region is dominated by central galaxies over a wide range of stellar masses. Below $10^{10}$ $M_{\odot}$, the relation is dominated by central galaxies of low mass halos with masses below $10^{12}-10^{12.5}$ $M_{\odot}$. Above the same stellar mass threshold, the MS region is populated mostly by star forming central galaxies of low and high mass groups from $10^{13}$ to $10^{14}$ $M_{\odot}$. The few star forming giant BCGs occupy the tail of the relation at high stellar masses. This distribution suggests a relation between the MS and the host halo mass, likely due to the tight correlation between the central galaxy stellar mass, $M_{*,c}$, and its host halo mass, $M_{halo}$ \citep[e.g.][]{moster10,behroozi13,yang17}. Indeed, \cite{popesso19a} show also that SFGs at the center of halos with mass larger than $10^{12}$ $M_{\odot}$ dominate numerically the MS in the region where the relation is bending. In this section we investigate whether there is a more general relation between the baryonic mass content of galaxies and their $M_{halo}$, able to explain the reduced availability of dust and molecular mass at the high mass end of the MS. 

We check this aspect by matching the star forming galaxy reference sample with the $M_{halo}$ catalog of \cite{yang05cat}. This catalog is based on SDSS DR7 and it provides a classification of galaxies in central/satellites and a measure of $M_{halo}$. A percentage of 93\% of the reference sample has a match in the \cite{yang05cat} catalog. Of this 89\% are classified as central galaxies. The halo mass is estimated through the correlation between the total group and cluster luminosity in the SDSS r band ($L_r$) and the halo mass. $L_r$ is estimated as the sum of the luminosities of the group members above a fixed absolute magnitude. The calibration of $L_r$ as a proxy for $M_{halo}$ in \cite{yang05cat} holds down to halo masses of $10^{11.5}$ $M_{\odot}$. To extend the halo mass info down to lower masses, namely for isolated galaxies with stellar mass below $10^{9.5}$ $M_{\odot}$, we estimate the mass of the host halo through the $M_{*,c}-M_{halo}$ relation of \cite{behroozi13}. To further check our results, we also use, as alternative halo mass estimate, the $M_{halo}$ derived by \cite{2014A&A...566A...1T}, based on the group velocity dispersion and under the assumption of a NFW profile.  

\begin{figure}
\includegraphics[width=\columnwidth]{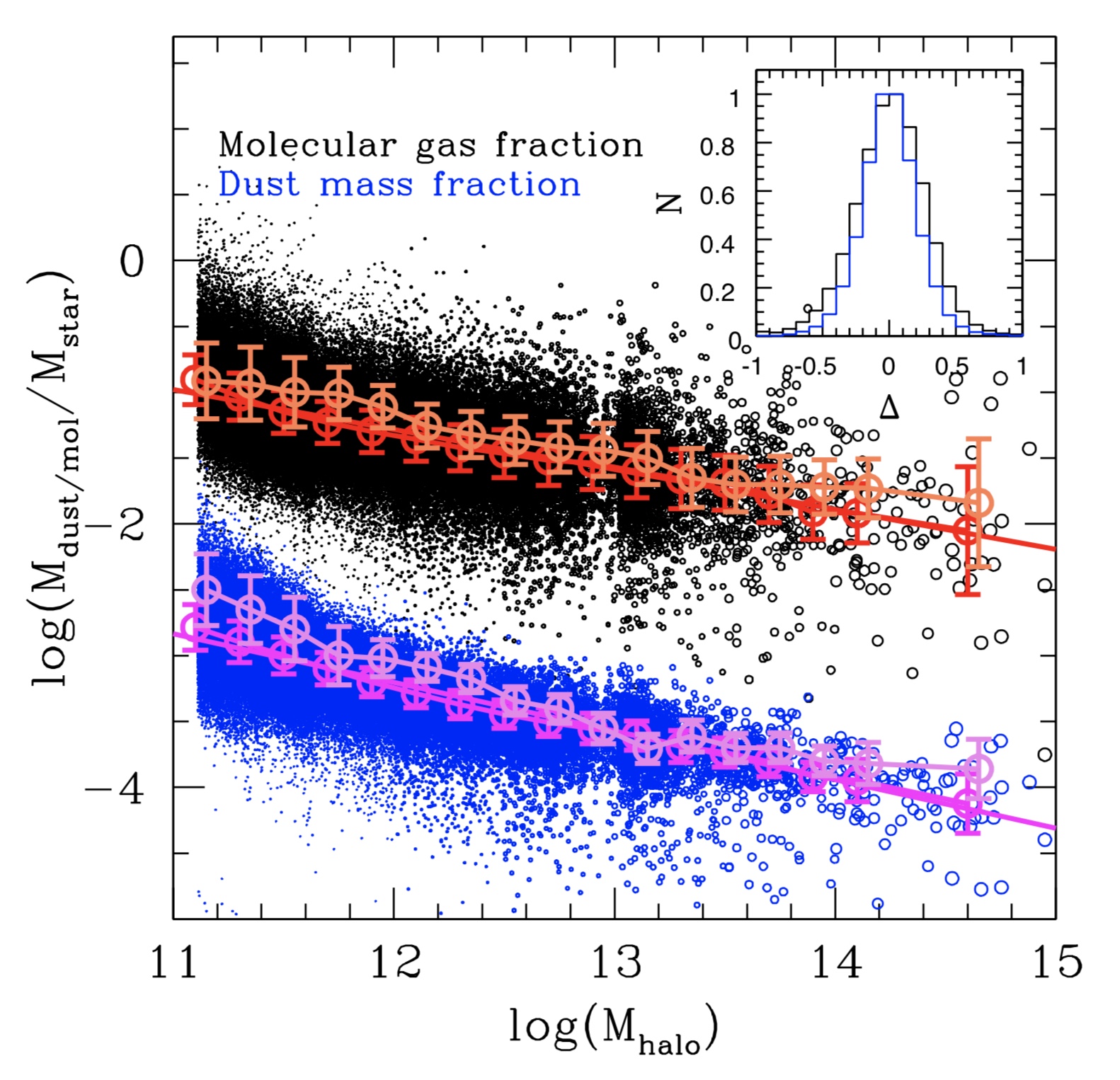}
\caption{ Dust (blue points) and molecular gas fractions (black points) as a function of $M_{halo}$ for all the central galaxies of the star forming galaxy reference sample (89\% of the sample).The magenta and red points show the median $f_{dust}-M_{halo}$ and $f_{mol}-M_{halo}$ relations, respectively, based on the halo mass estimates of Yang et al. (2007). The orange and violet points show the same relations, respectively, based on the halo masses of \protect\cite{2014A&A...566A...1T}.}
\label{f_star}
\end{figure}

\begin{figure}
\includegraphics[width=\columnwidth]{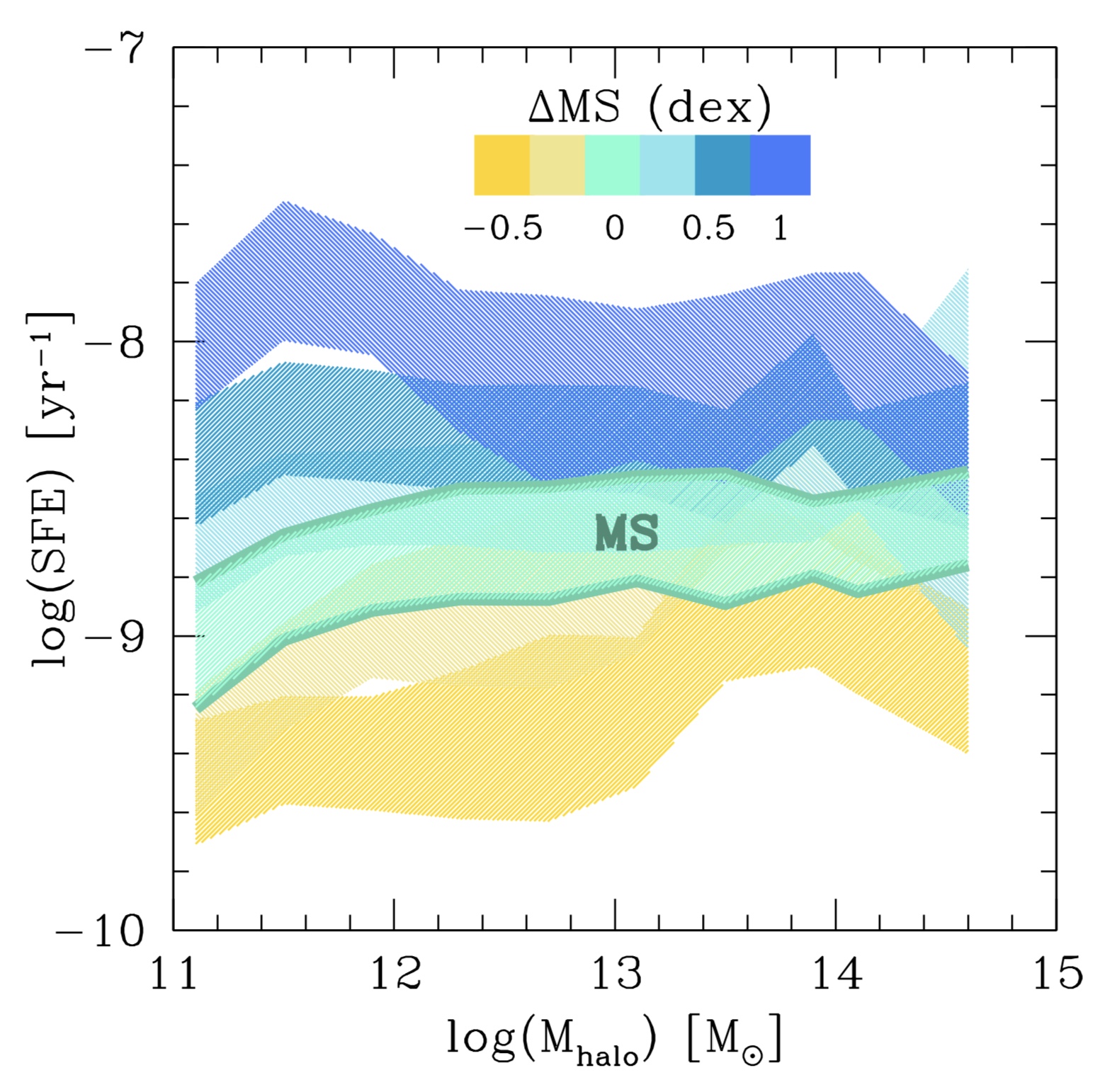}
\caption{Relation the star formation efficiency and the host halo mass. The different shaded regions indicate different bins of distance from the MS ($SFR/SFR_{MS}$), color-coded as indicated in the figure.}
\label{sfe_halo}
\end{figure}

Fig. \ref{f_star} shows how $f_{dust}$ (blue points) and $f_{mol}$ (black points) vary as a function of the $M_{halo}$. We find a very tight and significant anti-correlation for both quantities, which decrease by more than an order of magnitude from galaxies inhabiting $10^{11}$ $M_{\odot}$ halos to those in massive groups and clusters, with masses larger than $10^{14}$ $M_{\odot}$. The best fit power law is $f_{dust}\propto{M_{halo}}^{-0.36\pm0.02}$ and $f_{dust}\propto{M_{halo}}^{-0.31\pm0.03}$ The scatter is 0.19 and 0.21 dex for the $f_{dust}-M_{halo}$ and the $f_{mol}-M_{halo}$ relations, respectively. This would suggest that central galaxies of group and clusters are very inefficient in accreting or retaining their dust and molecular mass content with respect to galaxies at the center of lower mass halos. Interestingly, while $f_{dust}$ and $f_{mol}$ decrease rapidly with $M_{halo}$, the star formation efficiency does not show any dependence on $M_{halo}$, and it remains constant, at fixed distance from the MS, over the whole dynamic mass range considered here (see Fig. \ref{sfe_halo}). These results are not surprising, if we consider that $M_{halo}$ and $M_*$ are strongly correlated for central galaxies and that both $M_{dust}$ and $M_{mol}$ exhibit the strongest dependence on the stellar mass rather than the SFR (see Table 2).

\begin{figure}
\includegraphics[width=\columnwidth]{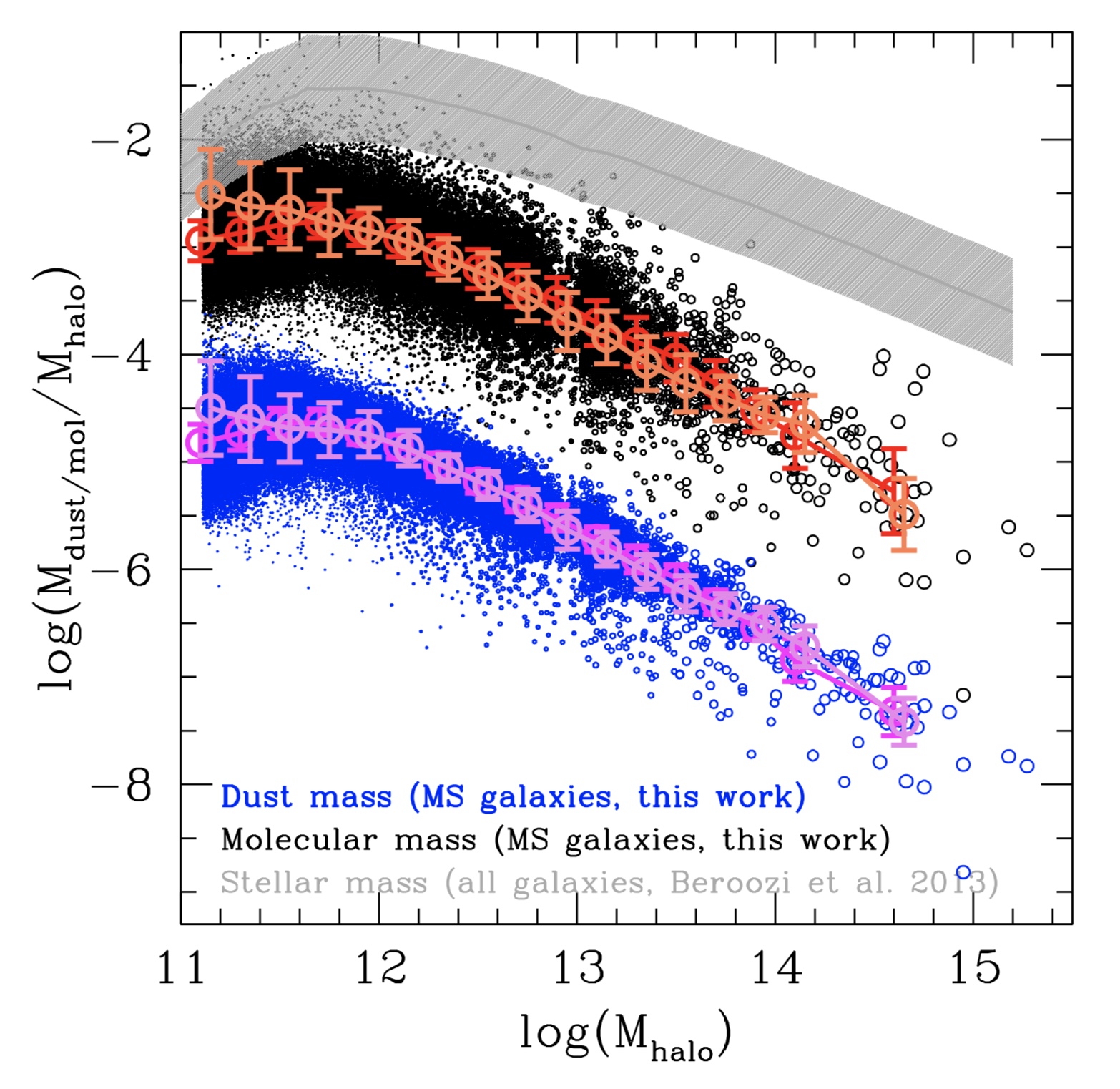}
\caption{Relation between the fraction of dust mass (blue points) and molecular mass (black points) over the host halo mass as a function of the host halo mass ($M_{halo}$) for the central galaxies of the star forming galaxy reference sample. The magenta and red connected empty points show the median $M_{dust}/M_{halo}-M_{halo}$ and $M_{mol}/M_{halo}-M_{halo}$ relations, based on the halo masses estimated by Yang et al. (2007). The orange and violet points show the same relations, respectively, based on the halo masses of \protect\cite{2014A&A...566A...1T}. The shaded region shows the $M_*/M_{halo}-M_{halo}$ relation of \protect\cite{behroozi13}.}
\label{f_halo}
\end{figure}

\begin{figure}
\includegraphics[width=\columnwidth]{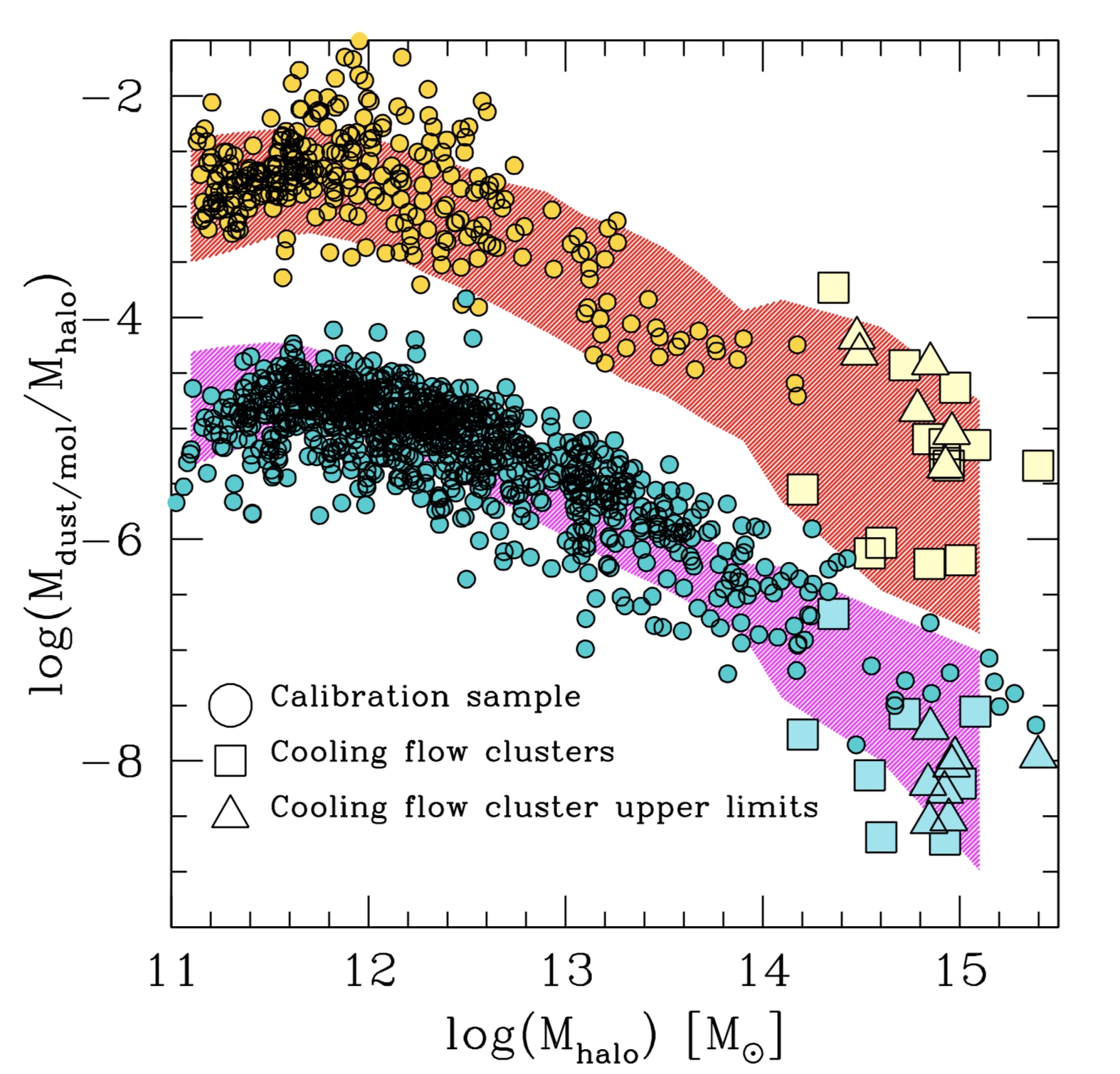}
\label{cool}
\caption{$M_{dust}/M_{halo}-M_{halo}$ (magenta region) and $M_{mol}/M_{halo}-M_{halo}$ (red region) relations as in Fig. \ref{f_halo}, with over-plotted the dust mass (cyan points) and molecular gas (yellow points) calibration samples, respectively. The dashed regions indicates the 3$\sigma$ regions of the relations. The squares indicate the dust (blue symbols) and molecular gas (yellow symbols) mass estimates of the cooling flow cluster sample of Edge et al. (2001). The triangles indicate the upper limits in the same sample, color coded as the squares.}
\label{cool}
\end{figure}

Similarly to what done in \cite{behroozi13} and \cite{behroozi19} for the efficiency of stellar mass growth, we estimate the efficiency in accreting/retaining the dust and molecular gas content as the fraction of dust and molecular gas over the host halo mass, for all central galaxies. Namely, we define the $M_{dust}/M_{halo}-M_{halo}$ and $M_{mol}/M_{halo}-M_{halo}$ relations in a similar way to the $M_{*,c}/M_{halo}-M_{halo}$ relation. These are shown in Fig. \ref{f_halo}. For comparison we plot also the the $M_{*,c}/M_{halo}-M_{Halo}$ relation of \citet[][gray shaded region]{behroozi13}. The shapes of the two relations resemble the $M_{*,c}/M_{halo}-M_{Halo}$ relation. Although our data do not go beyond halo masses of $10^{11}$ $M_{\odot}$, we can clearly see a peak at $10^{11.7}-10^{11.9}$ $M_{\odot}$ and a monotonically decreasing function towards larger masses, when using the \cite{yang05cat} catalog (red end magenta points in Fig. \ref{f_halo}). In this case the best power law fit at masses larger than $10^{12}$ $M_{\odot}$ is of the form $\propto{M_{halo}}^{-0.5}$ for both $M_{dust}/M_{halo}$ and $M_{mol}/M_{halo}$. Nevertheless, we point out that the position of the peak is very uncertain. Indeed, the halo mass estimates of \cite{2014A&A...566A...1T} lead to a consistent relation within 1$\sigma$ above $M_{halo}$ of $10^{12}$ $M_{\odot}$, but with a less evident peak. Below the peak towards lower $M_{halo}$, $M_{dust}/M_{halo}$ and $M_{mol}/M_{halo}$ tend to decrease. However, our stellar and halo mass limit and the uncertainty in the peak position prevents from robustly confirming the exact shape of the relation in this halo mass regime.  

 We underline that this relation is valid only for SFGs, where the proxies for $M_{dust}$ and $M_{mol}$ can be estimated. We speculate that the scatter of the relation should become larger in the regime of groups and clusters, which host predominantly inactive galaxies. 

To further check the reliability of such relations, we plot them in Fig. \ref{cool} as shaded regions, together with the dust and molecular mass calibration samples (cyan and yellow circles, respectively). In both cases the calibration samples lie in the region of the relations, although they under-sample the group and cluster mass regime. To check the relation in the cluster regime ($M_{halo} > 10^{14}$ $M_{\odot}$), we plot the $M_{dust}/M_{halo}$ and $M_{mol}/M_{halo}$ of the cooling flow cluster BCG sample of \cite{2001MNRAS.328..762E}. The sample is a collection of massive cooling flow systems with a central SFG, with dust and CO-based molecular mass estimate. The properties of the sample, including cluster name, redshift, $M_{halo}$,  $M_{dust}$ and $M_{mol}$ (or upper limits) are reported in Table \ref{cooling} with all the relative references (in the caption). All systems lie within the 3$\sigma$ regions of the $M_{dust}/M_{halo}-M_{halo}$ and $M_{mol}/M_{halo}-M_{halo}$ relations. The data are consistent with the shape, scatter and the slope of the relations also in the cluster regime.

We conclude that both dust and molecular gas contents of SFGs in the MS region are strictly related to the mass of the host halo. Thus, the relation between central galaxy and host halo does not affect only the stellar mass content but more in general the galaxy baryonic mass content. The progressive decrease of $M_{dust}/M_{halo}$ and $M_{mol}/M_{halo}$ towards the group and cluster dynamical range, indicate that such extreme environments are inhospitable sites to the SF process due to the lack of availability of molecular gas.

\begin{table*}
\begin{center}
\begin{tabular}{lcccc}
    \hline
    \hline \\[-2.5mm]
     Cluster & redshift & Mass &   $M_{dust}$ & $M_{mol}$\\
                  &              & ($10^{14}$ $M_{\odot}$) & ($M_{\odot}$) & ($M_{\odot}$) \\
  \hline
    \hline \\[-2.5mm]
A262			&	 0.0171		&  1.60$^a$	&	  4.4$\times 10^6$		 &	  9$\pm 1.3 \times 10^8$ \\
RXJ0352$+$19 	&	 0.109     		& 8.38$^a$      &	<6.9$\times 10^6$	 	 &	  1.2$\pm10.3 \times 10^{10}$\\
A478       		&	 0.0882  	        &10.10$^a$    	&	   1$\times 10^7$	  	 &	  1.3$\pm 1\times 10^9$	\\
RXJ0439$+$05 	&	 0.208    		&18.70$^a$	&  	<3.1$\times 10^7$	         &	<3.3$\times 10^{10}$	      \\ 	
A646       		&	 0.1268   		& 9.10$^a$	& 	<1.3$\times 10^7$	 	 &	<1.3$\times 10^{10}$         \\
RXJ0821$+$07 	&	 0.110      	& 5.25$^b$      &	  2.2$\times 10^7$	       	 &	3.9$\pm 0.4 \times 10^{10}$ \\
Zw2089     		&	 0.235    	  	& 16.6$^a$	&    	 <4.2$\times 10^7$	 &	<4.6$\times 10^{10}	$\\
Hydra-A                  &	 0.052     		&  4$^a$    	&	   1.3$\times 10^6$	 &	  7.4$\pm 1 \times 10^8$\\
Zw3146     		&	 0.2906  		& 30$^a$ 	&         2.2$\times 10^8$	 &	  1.6$\pm 0.3 \times 10^{11}$\\
A1068      		&	 0.1386   		&  2.29$^a$	&	   7.7$\times 10^7$ 	 &	  8.5$\times 10^{10}	$\\
Zw3916     		&	 0.204    		& 10.5$^c$	&	 <3.5$\times 10^8$ 	 &	 <1.6$\times 10^{10}$\\
A1664      		&	 0.1276   		&  9.46$^c$	&	 <1.6$\times 10^7$ 	 &	   4.4$\pm 0.7 \times 10^{10}  $\\
RXJ1347$-$11		&	 0.4503  		& 16$^d$	&	   1.8$\times 10^8$	 &	 <6.8$\times 10^{10}$ \\
A1795      		&	 0.0620   		&  7$^a$		&	 <3.1$\times 10^6$	 &          8.4$\pm 1 \times 10^8$\\
A1835      		&	 0.2523  		& 31.8$^a$	&	      1$\times 10^8$	 &          1.8$\pm 0.2 \times 10^{11}$\\
Zw7160     		&	 0.2578  		& 17.2$^a$	&	   1.5$\times 10^8$	 &	   6.1$\pm 2.4 \times 10^{10}$\\
RXJ1532$+$30 	&	 0.3615  		& 28.9$^a$	&	  <1.2$\times 10^8$	 &	   2.5$\pm 0.4\times 10^{11}	 $\\
A2146      		 &	 0.2343   	  	&  8.5$^e$ 	&	    5.6$\times 10^7$	 &       <3.5$\times 10^{10}$\\
A2204      		 &	 0.1514   		& 25$^a$		&	  <4.3$\times 10^7$	 &         2.3$\pm 0.6 \times 10^{10}$\\
Zw8193     		 &	 0.1825    	& 7.09$^f$	&	  <2.2$\times 10^7$	 &      <4.3$\times 10^{10}    $\\
Zw8197     		 &	 0.1140    	& 6.9$^f$	&	  <6.7$\times 10^6$	 &        1.1$\pm 0.3 \times 10^{10}$\\
Zw8276     		 &	 0.0757    	& 8.4$^a$	&	    2.6$\times 10^6$	 &        8.2$\pm 1.2 \times 10^9   $\\
A2390      		 &	 0.2328  	  	& 28$^a$	        &	       9$\times 10^7$	 &      <4.9$\times 10^{10}    $\\
A2597      		 &	 0.0852   		& 8.8$^f$	&	  <4.1$\times 10^6$	 &        8.1$\pm 3.3 \times 10^9$\\
NGC1275    	  	 &      0.0184   	& 12$^g$	&	    5.3$\times 10^7$	 &        1.7$\pm 0.2 \times 10^{10}  $\\
IRAS09104$+$4109 &      0.4420 		&     8.2$^a$    &           1.6$\times 10^8$	 &       <5.1$\times 10^{10}$\\
A1367      	     	 &      0.0218        	&  3.5$^a$	&	    4.1$\times 10^6$	 &           5.2$\pm 1 \times 10^8$\\
A2029			 &	 0.079		&  8.47$^a$     &          --                                 &	    <6$\times 10^9  $\\
A2199			 &	 0.03         	& 6.1$^a$        &         --     				 &      <1.4$\times 10^{10}  $\\
2A0335			 &	 0.0349     	& 3$^a$           &         --    				 &      <3.1$\times 10^{10}  $\\
A2052			 &	 0.03	 	& 3.1$^a$        &         --      				 &     <2.3$\times 10^{10}$\\

\hline
   \hline
\end{tabular}
\end{center}
\label{cooling}
\caption{The table lists the sample of cooling flow clusters, collected in Edge et al. (2001) with detection of dust and CO emission in the central galaxy. The dust and molecular mass estimates and upper limits are taken from the calibration of \protect\cite{2001MNRAS.328..762E}. All references for the individual detections can be found in that work. The cluster masses are estimated through different methods:
($^a$): cluster mass derived through Mass-temperature relation of \protect\cite{2001A&A...368..749F}. The temperature is derived from \protect\cite{1998MNRAS.301..881E, 2000MNRAS.318..333E}. The relation provide $M_{500}$, which is converted into $M_{200}$ assuming an NFW profile with a concentration parameter computed from the relation of \protect\cite{2007MNRAS.378...55M}; 
($^b$): cluster mass derived from \protect\cite{2007MNRAS.380...33H}; 
($^c$): cluster mass derived through X-ray luminosity ($L_X$)-mass relation of \protect\cite{2008MNRAS.387L..28R}.  $L_X$ is derived from from \protect\cite{2008MNRAS.384.1502S}; 
($^d$): cluster mass estimate of of \protect\cite{2010MNRAS.403.1787L}; 
($^e$): cluster mass measure of \protect\cite{2010MNRAS.405..257M}; 
($^f$): cluster mass derived through $L_X$-mass relation of \protect\cite{2008MNRAS.387L..28R}, with $L_X$ taken from \protect\cite{2017MNRAS.465.4872G}; 
($^g$): cluster mass derived from \protect\cite{1998MNRAS.300..837E}.}
\end{table*}

\section{Summary and Conclusions}

We summarize here the findings of our analysis.
We first investigate how the a combination of Balmer Decrement, disk inclination and gas metallicity correlates with the dust and molecular masses in a sample of star forming galaxies. We then use these proxies to estimate the dust and molecular masses for a large sample of local SDSS SFGs at $z< 0.1$. With these estimates available, we study the distribution of the dust and molecular mass along and across the Main Sequence relation of SGFs. 

The dust and molecular gas content can be expressed as a function of the SFR, or the distance from the MS, and the stellar mass. Both $M_{dust}$ and $M_{mol}$ tend to increase faster along the MS with the stellar mass than across the MS with the SFR or the distance from the MS. The dependence on the stellar mass is, in both cases, sub-linear. This implies that the fractions of dust and molecular mass content are decreasing along the MS, while they increase across the MS with the SFR. The different dependence on SFR and stellar mass implies also that the specific sSFR alone can not be a good proxy for $M_{dust}$ and $M_{mol}$ or their fractions. Similarly we find that the star formation efficiency (inverse of the depletion time) is nealy constant laong the MS and it depends marginally on the stellar mass and very significantly on the distance from the MS. SFE varies by more than 4 orders of magnitudes across the entire 3$\sigma$ MS region.

We also define the fraction of dust and molecular mass with respect to the halo mass as a measure of the halo efficiency in accreting or retaining their dust and molecular gas content. Such efficiency recalls the shape of the $M_{*,c}/M_{halo}-M_{Halo}$. The maximum is reached for central galaxies inhabiting halos of $10^{11.7}-10^{11.9}$ $M_{\odot}$ and it decreases by more than two orders of magnitude towards group- and cluster-sized halos. Conversely, the star formation efficiency is constant as a function of the halo mass and depends only on the distance from the MS. As the region where the MS is bending is numerically dominated by such massive halos, we conclude that the bending is due to a lower availability of baryonic mass in massive halos rather than a lower efficiency in forming stars.

\section*{Acknowledgements}
This research was supported by the DFG cluster of excellence "Origin and Structure of the Universe" (www.universe-cluster.de).

\section*{Data availability}
The data underlying this article are available at the SDSS database (https://www.sdss.org/dr16/), the MPA-JHU database of galaxy  properties (https://www.mpa.mpa-garching.mpg.de/SDSS/DR7/), the CO Legacy Database for GASS survey (http://www.star.ucl.ac.uk/xCOLDGASS/data.html), the Herschel SPIRE-selected sample of Bertemes et al. (2018), the H-ATLAS DR2 (https://www.h-atlas.org/public-data/download) and the Herschel Reference Survey (https://hedam.lam.fr/HRS/).



\bibliographystyle{mnras}
\bibliography{mdust} 




\bsp	
\label{lastpage}
\end{document}